\documentclass{article}

\PassOptionsToPackage{square,numbers,sort&compress}{natbib}

\usepackage[preprint]{neurips_2020}

\usepackage[utf8]{inputenc} 
\usepackage[T1]{fontenc}    
\usepackage[hidelinks]{hyperref}       
\usepackage{url}            
\usepackage{booktabs}       
\usepackage{amsfonts}       
\usepackage{nicefrac}       
\usepackage{microtype}      

\usepackage{amsmath,mathtools,amsthm}
\usepackage{color}

\usepackage[ruled,linesnumbered]{algorithm2e}
\usepackage{algorithmicx}
\usepackage{booktabs}
\usepackage{subfigure}

\usepackage{xr}
\makeatletter
\newcommand*{\addFileDependency}[1]{
  \typeout{(#1)}
  \@addtofilelist{#1}
  \IfFileExists{#1}{}{\typeout{No file #1.}}
}
\makeatother

\newcommand*{\myexternaldocument}[1]{
    \externaldocument{#1}
    \addFileDependency{#1.tex}
    \addFileDependency{#1.aux}
}
\myexternaldocument{supplement_nips}

\usepackage{fancyhdr}
\usepackage{pdfpages}
\newcommand\invisiblesection[1]{%
  \refstepcounter{section}%
  \addcontentsline{toc}{section}{\protect\numberline{\thesection}#1}%
  \sectionmark{#1}}

\newcommand{\ceil}[1]{\left\lceil#1\right\rceil}
\newtheorem{theorem}{Theorem}

\title{Classification with Valid and Adaptive Coverage}

\author{%
  Yaniv Romano\thanks{Equal contribution.} \\
  Department of Statistics\\
  Stanford University\\
  \And
  Matteo Sesia\footnotemark[1] \\
  Department of Statistics \\
  Stanford University \\
  \And
  Emmanuel J. Cand{\`e}s \\
  Departments of Mathematics \\
  and of Statistics \\
  Stanford University \\
}

\begin{document}

\maketitle

\begin{abstract}
Conformal inference, cross-validation+, and the jackknife+ are hold-out methods that can be combined with virtually any machine learning algorithm to construct prediction sets with guaranteed marginal coverage. In this paper, we develop specialized versions of these techniques for categorical and unordered response labels that, in addition to providing marginal coverage, are also fully adaptive to complex data distributions, in the sense that they perform favorably in terms of approximate conditional coverage compared to alternative methods.
The heart of our contribution is a novel conformity score, which we explicitly demonstrate to be powerful and intuitive for classification problems, but whose underlying principle is potentially far more general.
Experiments on synthetic and real data demonstrate the practical value of our theoretical guarantees, as well as the statistical advantages of the proposed methods over the existing alternatives. 
\end{abstract}

\section{Introduction}

Imagine we have $n$ data samples $\{ (X_i, Y_i)\}_{i=1}^n$ with features $X_{i} \in \mathbb{R}^{p}$ and a discrete label $Y_{i}\in \mathcal{Y} = \{1,2,\dots,C\}$. The samples are drawn exchangeably (e.g., i.i.d., although independence is unnecessary) from some unknown distribution $P_{XY}$. Given such data and a desired coverage level $1-\alpha \in (0,1)$, we seek to construct a prediction set $\hat{\mathcal{C}}_{n,\alpha} \subseteq \mathcal{Y}$ for the {\em unseen} label of a new data point $(X_{n+1}, Y_{n+1})$, also drawn exchangeably from $P_{XY}$, achieving marginal coverage; that is, obeying 
\begin{align} \label{eq:marginal-coverage}
    \mathbb{P}\left[Y_{n+1} \in \hat{\mathcal{C}}_{n,\alpha}(X_{n+1}) \right] \geq 1-\alpha.
\end{align}
 The probability above is taken over all $n+1$ data points, and we ask that \eqref{eq:marginal-coverage} holds for any fixed $\alpha$, $n$, and $P_{XY}$. While marginal coverage has the advantage of being both desirable and practically achievable, 
it unfortunately does not imply the stronger notion of conditional coverage:
\begin{align} \label{eq:conditional-coverage}
    \mathbb{P}\left[ Y_{n+1} \in \hat{\mathcal{C}}_{n,\alpha}(x) \mid X_{n+1} = x \right] \geq 1-\alpha.
\end{align}
The latter asks for valid coverage conditional on a specific observed value of the features~$X$.
It is already known that conditional coverage cannot be achieved in theory without strong modeling assumptions~\cite{vovk2012conditional, barber2019limits}, which we are not willing to make in this paper. That said, it is undeniable that conditional coverage would be preferable.
We thus seek to develop classification methods that are provably valid in the marginal sense~\eqref{eq:marginal-coverage} and also attempt to sensibly approximate conditional coverage~\eqref{eq:conditional-coverage}. At the same time, we want powerful predictions, in the sense that the cardinality of $\hat{\mathcal{C}}$ should be as small as possible.

\subsection{The oracle classifier} \label{sec:oracle}

Imagine we have an \textit{oracle} with perfect knowledge of the conditional distribution $P_{Y \mid X}$ of $Y$ given $X$.
This would of course give the problem away; to be sure, we would define optimal prediction sets $\mathcal{C}^{\text{oracle}}_{\alpha}(X_{n+1})$ with conditional coverage as follows: 
for any $x \in \mathbb{R}^p$, set $\pi_{y}(x) = \mathbb{P}[ Y = y \mid X =x]$ for each $y \in \mathcal{Y}$.
Denote by $\pi_{(1)}(x) \ge \pi_{(2)}(x) \geq \ldots \geq \pi_{(C)}(x)$ the order statistics for $\pi_{y}(x)$.
For simplicity, let us assume for now that there are no ties; we will relax this assumption shortly.
For any $\tau \in [0,1]$, define the \emph{generalized conditional quantile} function\footnote{Recall that the conditional quantiles for continuous responses are: $\inf \{y \in \mathbb{R} : \mathbb{P}[Y \leq y \mid X=x] \geq \tau\}$.}
\begin{align} \label{eq:oracle-threshold}
  L(x; \pi, \tau) & = \min \{ c \in \{1,\ldots,C\} \ : \ \pi_{(1)}(x) + \pi_{(2)}(x) + \ldots + \pi_{(c)}(x) \geq \tau \},
  \end{align}
and the prediction set:
\begin{align} \label{eq:oracle-sets-det}
  C^{\text{oracle+}}_{\alpha}(x) & = \left\{ \text{`$y$' indices of the $L(x; \pi, 1-\alpha)$ largest } \pi_{y}(x) \right\}.
\end{align}
Hence, \eqref{eq:oracle-sets-det} is the smallest deterministic set that contains a response with feature values $X = x$ with probability at least $1-\alpha$. For example, if $\pi_{1}(x) = 0.3$, $\pi_{2}(x) = 0.6$, and $\pi_{3}(x)=0.1$, we have $\pi_{(1)}(x) = 0.6$, $\pi_{(2)}(x) = 0.3$, and $\pi_{(3)}(x)=0.1$, with $L(x,0.9)=2$, $C^{\text{oracle}}_{0.1}(x) = \{1,2\}$, and $L(x,0.5)=1$, $C^{\text{oracle}}_{0.5}(x) = \{2\}$.
Furthermore, define a function $\mathcal{S}$ with input $x$, $u \in [0,1]$, $\pi$, and $\tau$, which can be seen as a \emph{generalized inverse} of~\eqref{eq:oracle-threshold}:
\begin{align} \label{eq:define-S}
    \mathcal{S}(x, u ; \pi, \tau) & = 
    \begin{cases}
    \text{ `$y$' indices of the $L(x ; \pi,\tau)-1$ largest $\pi_{y}(x)$},
    & \text{ if } u \leq V(x ; \pi,\tau) , \\
    \text{ `$y$' indices of the $L(x ; \pi,\tau)$ largest $\pi_{y}(x)$},
    & \text{ otherwise},
    \end{cases}
\end{align}
where
\begin{align*}
    V(x; \pi, \tau) & =  \frac{1}{\pi_{(L(x ; \pi, \tau))}(x)} \left[\sum_{c=1}^{L(x ; \pi, \tau)} \pi_{(c)}(x) - \tau \right].
\end{align*}
With this in place, by letting $u$ be the realization of a uniform random variable, we can see that the oracle has access to tighter randomized prediction sets, namely,
\begin{align} \label{eq:oracle-sets}
  C^{\text{oracle}}_{\alpha}(x) = \mathcal{S}(x, U; \pi, 1-\alpha).
\end{align}
Above, $U \sim \text{Uniform}(0,1)$ is independent of everything else.
It is easy to verify that the sets in~\eqref{eq:oracle-sets} are the smallest randomized prediction sets with conditional coverage at level $1-\alpha$. 
In the above example, we would have $C^{\text{oracle}}_{0.5}(x)=\emptyset$ with probability $(0.6-0.5)/0.6=1/6$ and $C^{\text{oracle}}_{0.5}(x) = \{2\}$ otherwise.
Finally, if there are any ties among the class probabilities, the oracle could simply break them at random and discard from $C^{\text{oracle}}_{\alpha}(x)$ all labels with zero probability.
Of course, we do not have access to such an oracle since $P_{Y \mid X}$ is unknown.

\subsection{Preview of our methods}

This paper uses classifiers trained 
on the available data to approximate the unknown conditional distribution of $Y \mid X$. 
A key strength of the proposed methods is their ability to work with any black-box predictive model, including neural networks, random forests, support vector classifiers, or any other currently existing or possible future alternatives. The only restriction on the training algorithm is that it should treat all samples exchangeably; i.e., it should be invariant to their order. 
Most off-the-shelf tools offer such suitable probability estimates $\hat{\pi}_{y}(x)$ that we can exploit, regardless of whether they are well-calibrated, by imputing them into an algorithm inspired by the oracle from Section~\ref{sec:oracle} in order to obtain prediction sets with guaranteed coverage---as we shall see.

Our reader will understand that naively substituting ${\pi}_{y}(x)$ with $\hat{\pi}_{y}(x)$ into the oracle procedure would yield predictions lacking any statistical guarantees because $\hat{\pi}_{y}(x)$ may be a poor approximation of  $\pi_{y}(x)$.
Fortunately, we can automatically account for errors in $\hat{\pi}_{y}(x)$ by adaptively choosing the threshold $\tau$ in~\eqref{eq:oracle-threshold} in such a way as to guarantee finite-sample coverage on future test points.

\subsection{Related work}

We build upon conformal inference \cite{vovk2005algorithmic,vovk2009line,lei2018distribution} and take inspiration from \cite{lei2014distribution,chernozhukov2019distributional, kivaranovic2019adaptive, kuchibhotla2019nested,guan2019conformal,izbicki2019distribution,romano2019conformalized} which made conformal prediction for regression problems adaptive to heteroscedasticity, thus bringing it closer to conditional coverage~\cite{sesia2019comparison}.
Conformal inference has been applied before to classification problems \cite{vovk2003mondrian,vovk2005algorithmic,hechtlinger2018cautious,sadinle2019least} in order to attain marginal coverage; however, the idea of explicitly trying to approximate the oracle from  Section~\ref{sec:oracle} is novel. We will see that our procedure empirically achieves  better conditional coverage than a direct application of conformal inference. 
While working on this project, we became aware of the independent work of~\cite{cauchois2020knowing}, which also seeks to improve the conditional coverage of conformal classification methods.
However, their approach differs substantially; see Section~\ref{sec:methods-others}.
Finally, our method also naturally accommodates calibration through cross-validation+ and the jackknife+~\cite{barber2019predictive}, which had not yet been extended to classification, although the natural generality of these calibration techniques has also been very recently noted by others \cite{kuchibhotla2019nested}.

A different but related line of work focuses on post-processing the output of black-box classification algorithms to produce more accurate probability estimates~\cite{platt1999probabilistic,zadrozny2001obtaining,zadrozny2002transforming, vaicenavicius2019evaluating,guo2017calibration,neumann2018relaxed,kumar2019verified}, although without achieving prediction sets with provable finite-sample coverage. These techniques are complementary to our methods and may help further boost our performance by improving the accuracy of any given black box; however, we have not tested them empirically in this paper for space reasons. 

\section{Methods}

\subsection{Generalized inverse quantile conformity scores} \label{sec:methods-scores}


Suppose we have a black-box classifier $\hat{\pi}_{y}(x)$ that estimates the true unknown class probabilities ${\pi}_{y}(x)$. Here, we only assume $\hat{\pi}_{y}(x)$ to be standardized:
$0 \leq \hat{\pi}_{y}\left(x\right) \leq 1$,
$\sum_{y=1}^{C} \hat{\pi}_{y}\left(x \right) = 1$, $\forall x,y$.
An example may be the output of the softmax layer of a neural network, after normalization. In fact, almost any standard machine learning software, e.g., \texttt{sklearn}, can produce a suitable $\hat{\pi}$, either through random forests, k-nearest neighbors, or support vector machines, to name a few options.
Then, we plug $\hat{\pi}$ into a modified version of the imaginary oracle procedure of Section~\ref{sec:oracle} where the threshold $\tau$ needs to be carefully calibrated using hold-out samples independent of the training data.
We will present two alternative methods for calibrating $\tau$; both are based on the following idea.

Define a \textit{generalized inverse quantile} conformity score function $E$ with input $x,y,u,\hat{\pi}$,
\begin{align} \label{eq:define-scores}
    E(x,y,u;\hat{\pi}) & = \min \left\{ \tau \in [0,1] : y \in \mathcal{S}(x, u ; \hat{\pi}, \tau) \right\},
\end{align}
where $\mathcal{S}$ is our generalized notion of (estimated) conditional quantiles, defined in~\eqref{eq:define-S}.
The conformity score $E(\cdot)$ is the inverse of the smallest generalized quantile that contains the label $y$ conditional on $X=x$. 
{By construction, our scores evaluated on hold-out samples $(X_i,Y_i)$, namely $E_i = E(X_i,Y_i,U_i; \hat{\pi})$, are uniformly distributed conditional on $X$ if $\hat{\pi} = \pi$. (Each $U_i$ is a uniform random variable in $[0,1]$ independent of everything else.) Therefore, one could also intuitively look at~\eqref{eq:define-scores} as a special type of \textit{p-value}.
It is worth emphasizing that this property makes our scores naturally comparable across different samples, in contrast with the scores found in the earlier literature on adaptive conformal inference \cite{romano2019conformalized}. In fact, alternative conformity scores \cite{lei2018distribution,romano2019conformalized,kuchibhotla2019nested,cauchois2020knowing} generally have different distributions at different values of $X$, even in the ideal case where the base method (our $\hat{\pi})$ is a perfect oracle.
Below, we shall see that, loosely speaking, we can construct prediction sets with provable marginal coverage for future test points by applying~\eqref{eq:define-S} with a value of $\tau$ close to the $1-\alpha$ quantile of $\{E_i\}_{i \in \mathcal{I}_2}$, where $\mathcal{I}_2$ is the set of hold-out data points not used to train~$\hat{\pi}$.
}

\subsection{Adaptive classification with split-conformal calibration} \label{sec:methods-split}

Algorithm~\ref{alg:sc} implements the above idea with split-conformal calibration, from which we begin because it is the easiest to explain.
Later, we will consider alternative calibration methods based on cross-validation+ and the jackknife+; we do not discuss full-conformal calibration in the interest of space, and because it is often computationally prohibitive.
For simplicity, we will apply Algorithm~\ref{alg:sc} by splitting the data into two sets of equal size; however, this is not necessary and using more data points for training may sometimes perform better in practice~\cite{sesia2019comparison}. 

\begin{algorithm}[!htb]
  \SetAlgoLined
    \textbf{Input:} data $\left\{(X_i, Y_i)\right\}_{i=1}^{n}$, $X_{n+1}$, black-box learning algorithm $\mathcal{B}$, level $\alpha \in (0,1)$. \\
  Randomly split the training data into $2$ subsets, $\mathcal{I}_1,\mathcal{I}_2$. \\
    Sample $U_i \sim \text{Uniform}(0,1)$ for each $i \in \{1,\ldots,n+1\}$, independently of everything else. \\
  Train $\mathcal{B}$ on all samples in $\mathcal{I}_1$:
  $\hat{\pi} \leftarrow \mathcal{B}( \{(X_i,Y_i)\}_{i\in \mathcal{I}_1} )$. \\
   Compute $E_i  = E(X_i, Y_i, U_i; \hat{\pi})$ for each $i \in \mathcal{I}_2$, with the function $E$ defined in~\eqref{eq:define-scores}.\\
  Compute $\hat{Q}_{1-\alpha}(\{E_i \}_{i \in \mathcal{I}_2})$ as the $\ceil{(1-\alpha)(1+|\mathcal{I}_2|)}$th largest value in $\{E_i \}_{i \in \mathcal{I}_2}$. \\
  Use the function $\mathcal{S}$ defined in~\eqref{eq:define-S} to construct the prediction set at $X_{n+1}$ as:
  \begin{align} \label{eq:predictions-SC}
    \hat{\mathcal{C}}^{\mathrm{sc}}_{n,\alpha}(X_{n+1}) = \mathcal{S}(X_{n+1}, U_{n+1} ;\hat{\pi},  \hat{Q}_{1-\alpha}(\{E_i \}_{i \in \mathcal{I}_2})).
  \end{align}\\
  \textbf{Output:} A prediction set $\hat{\mathcal{C}}^{\text{SC}}_{n,\alpha}(X_{n+1})$ for the unobserved label $Y_{n+1}$.
 \caption{Adaptive classification with split-conformal calibration}
 \label{alg:sc}
\end{algorithm}

\begin{theorem} \label{thm:SC}
If the samples $(X_i,Y_i)$, for $i \in \{1,\dots,n+1\}$, are exchangeable and $\mathcal{B}$ from Algorithm~\ref{alg:sc} is invariant to permutations of its input samples, the output of~Algorithm~\ref{alg:sc} satisfies:
\begin{align}
  \mathbb{P}\left[Y_{n+1} \in \hat{\mathcal{C}}^{\mathrm{sc}}_{n,\alpha}(X_{n+1}) \right] \geq 1-\alpha.
\end{align}
Furthermore, if the scores $E_i$ are almost surely distinct, the marginal coverage is near tight:
\begin{align}
  \mathbb{P}\left[Y_{n+1} \in \hat{\mathcal{C}}^{\mathrm{sc}}_{n,\alpha}(X_{n+1})\right] \leq 1-\alpha + \frac{1}{|\mathcal{I}_2|+1}.
\end{align}
\end{theorem}

The proofs of this theorem and all other results are in Supplementary Section~\ref*{sec:supp-proofs}.
Marginal coverage holds regardless of the quality of the black-box approximation;
however, one can intuitively expect that if the black-box is consistent and a large amount of data is available, so that $\hat{\pi}_{y}(x) \approx \pi_{y}(x)$, the output of our procedure will tend to be a close approximation of the output of the oracle, which provides optimal conditional coverage.
This statement could be made rigorous under some additional technical assumptions besides the consistency of the black box~\cite{sesia2019comparison}.
However, we prefer to avoid tedious technical details, especially since the intuition is already clear. If $\hat{\pi} = \pi$, the sets $\mathcal{S}(X_i,U_i ; \pi, \tau)$ in~\eqref{eq:define-S} will tend to contain the true labels for a fraction $\tau$ of the points $i \in \mathcal{I}_2$, as long as $|\mathcal{I}_2|$ is large. In this limit, $\hat{Q}_{1-\alpha}(\{E_i \}_{i \in \mathcal{I}_2})$ becomes approximately equal to $1-\alpha$, and the predictions in~\eqref{eq:predictions-SC} will eventually approach those in~\eqref{eq:oracle-sets}.

\subsection{Adaptive classification with cross-validation+ and jackknife+ calibration}  \label{sec:methods-cv+}

A limitation of Algorithm~\ref{alg:sc} is that it only uses part of the data to train the predictive algorithm. Consequently, the estimate $\hat{\pi}$ may not be as accurate as it could have been had we used all the data for estimation purposes. This is especially true if the sample size $n$ is small. 
Algorithm~\ref{alg:cv+} presents an alternative solution that replaces data splitting with a cross-validation approach, which is  computationally more expensive but often provides tighter prediction sets.

\begin{algorithm}[!htb]
  \SetAlgoLined
  \textbf{Input:} data $\left\{(X_i, Y_i)\right\}_{i=1}^{n}$, $X_{n+1}$, black-box $\mathcal{B}$, number of splits $K \leq n$, level $\alpha \in (0,1)$. \\
  Randomly split the training data into $K$ disjoint subsets, $\mathcal{I}_1,\dots,\mathcal{I}_K$, each of size $n/K$. \\
  Sample $U_i \sim \text{Uniform}(0,1)$ for each $i \in \{1,\ldots,n+1\}$, independently of everything else. \\
  \For{$k \in \{1, \ldots, K\}$} {
    Train $\mathcal{B}$ on all samples except those in $\mathcal{I}_k$:
    $\hat{\pi}^{k} \leftarrow \mathcal{B}( \{(X_i,Y_i)\}_{i\in \{1,\dots,n\} \setminus \mathcal{I}_k} )$. \\
    }
  Use the function $E$ defined in~\eqref{eq:define-scores} to construct the prediction set $\hat{\mathcal{C}}^{\text{CV+}}_{n,\alpha}(X_{n+1})$ as:
  \begin{align} \label{eq:predictions-CV+}
  \begin{split}
    &\hat{\mathcal{C}}^{\text{CV+}}_{n,\alpha}(X_{n+1}) =
    \bigg\{ y \in \mathcal{Y}:  \\ & \quad \quad \sum_{i=1}^n \mathbf{1}\left[ E(X_i,Y_i,U_i; \hat{\pi}^{k(i)}) < E(X_{n+1},y,U_{n+1}; \hat{\pi}^{k(i)}) \right] < (1-\alpha)(n+1) \bigg\},
    \end{split}
  \end{align}
  where $k(i) \in \{1,\ldots,K\}$ is the fold containing the $i$th sample. \\
  \textbf{Output:} A prediction set $\hat{\mathcal{C}}^{\text{CV+}}_{n,\alpha}(X_{n+1})$ for the unobserved label $Y_{n+1}$.
 \caption{Adaptive classification with CV+ calibration}
 \label{alg:cv+}
\end{algorithm}

In words, in Algorithm~\ref{alg:cv+}, we sweep over all possible labels $y \in \mathcal{Y}$ and include $y$ in the final prediction set $\hat{\mathcal{C}}^{\text{CV+}}_{n,\alpha}(X_{n+1})$ if the corresponding score $E(X_{n+1},y,{U_{n+1}}; \hat{\pi}^{k(i)})$ is smaller than $(1-\alpha)(n+1)$ hold-out scores $E(X_{i},Y_i,{U_{i}}; \hat{\pi}^{k(i)})$ evaluated on the true labeled data.
Note that we have assumed $n/K$ to be an integer for simplicity; however, different splits can have different sizes. In the special case where $K=n$, we refer to the hold-out system in Algorithm~\ref{alg:cv+} as jackknife+ rather than cross-validation+, consistently with the terminology in~\cite{barber2019predictive}.

\begin{theorem} \label{thm:CV+}
Under the same assumptions of Theorem~\ref{thm:SC}, the output of~Algorithm~\ref{alg:cv+} satisfies:
\begin{align}
  \mathbb{P} \left[ \{Y_{n+1} \in \hat{\mathcal{C}}^{\mathrm{CV+}}_{n,\alpha}(X_{n+1}) \right] \geq 1-2\alpha - \min \left\{\frac{2(1-1/K)}{n/K+1}, \frac{1-K/n}{K+1} \right\}.
\end{align}
In the special case where $K=n$, this bound simplifies to:
\begin{align}
  \mathbb{P} \left[ Y_{n+1} \in \hat{\mathcal{C}}^{\mathrm{JK+}}_{n,\alpha}(X_{n+1}) \right] \geq 1-2\alpha.
\end{align}
\end{theorem}

Note that this establishes that the coverage is slightly below $1-2\alpha$.
Therefore, to guarantee $1-\alpha$ coverage, we should replace the input $\alpha$ in Algorithm~\ref{alg:cv+} with a smaller value near $\alpha/2$.
We chose not to do so because our experiments show that the current implementation already typically covers at level $1-\alpha$ (or even higher) in practice; this empirical observation is consistent with~\cite{barber2019predictive}.
Furthermore, there exists a conservative variation of Algorithm~\ref{alg:cv+} for which we can prove $1-\alpha$ coverage without modifying the input level; see Supplementary Section~\ref*{sec:supp-methods-minimax-JK+}.

To see why everything above makes sense, consider what would happen if the black-box estimates of conditional probabilities in Algorithm~\ref{alg:cv+} were exact.
In this case, the final prediction set in~\eqref{eq:predictions-CV+} would become
  \begin{align}
  \begin{split}
    \hat{\mathcal{C}}^{\text{CV+}}_{n,\alpha}(X_{n+1})
    & = \left\{ y \in \mathcal{Y}: E(X_{n+1},y,U_{n+1}; \pi) < \hat{Q}_{1-\alpha}(
    \{E(X_i,Y_i,U_i; \pi)\}_{i \in \{1,\ldots,n\}}) \right\},
    \end{split}
  \end{align}
where $\hat{Q}_{1-\alpha}$ is defined as in Section~\ref{sec:methods-split}.
If $n$ is large, for any fixed threshold $\tau$, we can expect $\mathcal{S}(X_i, U_i; \pi, \tau)$ to contain $Y_i$ for approximately a fraction $\tau$ of samples $i$. Therefore, $\hat{Q}_{1-\alpha}(\{E(X_i,Y_i,U_i; \pi)\}_{i \in \{1,\ldots,n\}}) \approx 1-\alpha$, and the decision rule becomes approximately:
\begin{align}
  \hat{\mathcal{C}}^{\text{CV+}}_{n,\alpha}(X_{n+1})
  \approx \left\{ y \in \mathcal{Y}:
  E(X_{n+1},y,U_{n+1}; \pi) \leq 1-\alpha \right\},
\end{align}
which is equivalent to the oracle procedure from Section~\ref{sec:oracle}.

\subsection{Comparison with alternative conformal methods} \label{sec:methods-others}

Conformal prediction has been proposed before in the context of classification~\cite{vovk2005algorithmic}, through a very general calibration rule of the form
\begin{align*}
  \hat{\mathcal{C}}(x; t) = \{ y \in \mathcal{Y} : \hat{f}(y \mid x) \geq t \},
\end{align*}
where the score $\hat{f}$ is a function learned by a black-box classifier. 
However, to date it was not clear how to best translate the output of the classifier into a powerful score $\hat{f}$ for the above decision rule.
In fact, typical choices of $\hat{f}(y \mid x)$, e.g., the estimated probability of $Y=y$ given $X=x$,
often lead to poor conditional coverage because the same threshold $t$ is applied both to easy-to-classify samples (where one label has probability close to 1 given $X$) and to hard-to-classify samples (where all probabilities are close to $1/|\mathcal{Y}|$ given $X$). Therefore, this \textit{homogeneous} conformal classification may significantly underperform compared to the oracle from Section~\ref{sec:oracle}, even in the ideal case where the black-box manages to learn the correct probabilities.
This limitation has also been very recently noted in~\cite{cauchois2020knowing} and is analogous to that addressed by~\cite{romano2019conformalized} in problems with a continuous response variable~\cite{sesia2019comparison}.

The work of~\cite{cauchois2020knowing} addresses this problem by applying quantile regression~\cite{romano2019conformalized} to hold-out scores $\hat{f}$. However, their solution has two limitations. Firstly, it involves additional data splitting to avoid overfitting, which tends to reduce power. Secondly, its theoretical asymptotic optimality is weaker than ours because it requires the consistency of two black-boxes instead of one (this should be clear even though we have explained consistency only heuristically).
Practically, experiments suggest that our method provides superior conditional coverage and often yields smaller prediction sets.

\section{Experiments with simulated data} \label{sec:simulated}

\subsection{Methods and metrics}

We compare the performances of Algorithms~\ref{alg:sc} (SC) and~\ref{alg:cv+} (CV+, JK+), which are based on the new generalized inverse quantile conformity scores in~\eqref{eq:define-scores}, to those of homogeneous conformal classification (HCC) and conformal quantile classification (CQC)~\cite{cauchois2020knowing}.
We focus on two different data generating scenarios in which marginal coverage is not a good proxy for conditional coverage (the second setting is discussed in Supplementary Section~\ref*{sec:supp-experiments-2}).
In both cases, we explore 3 different black-boxes:
an \textit{oracle} that knows the true 
$\pi_{y}(x)$ for all $y \in \mathcal{Y}$ and $x$;
a support vector classifier (SVC) implemented by the \texttt{sklearn} Python package;
and a random forest classifier (RFC) also implemented by \texttt{sklearn}---
see Supplementary Section~\ref*{sec:supp-bb-details} for more details.

We fix $\alpha=0.1$ and assess performance in terms of marginal coverage, conditional coverage, and mean cardinality of the prediction sets.
Conditional coverage is defined using an estimate of the worst-slice (WS) coverage similar to that in~\cite{cauchois2020knowing}, as explained in Supplementary Section~\ref*{sec:supp-cond-coverage}.
The cardinality of the prediction sets is computed both marginally and conditionally on coverage; the former is defined as $\mathbb{E}[|\hat{\mathcal{C}}(X_{n+1})|]$ and the latter as $\mathbb{E}[|\hat{\mathcal{C}}(X_{n+1})| \mid Y_{n+1} \in \hat{\mathcal{C}}(X_{n+1})]$.
Additional coverage and size metrics defined by conditioning on the value of a given discrete feature, e.g., $X_1$, are discussed in Supplementary Section~\ref*{sec:supp-experiments-sim}.

\subsection{Experiments with multinomial model and inhomogeneous features}
\label{sec:experiments-1}

We generate the features $X \in \mathbb{R}^{p}$, with $p=10$, as follows:
$X_1=1$ w.p. $1/5$ and $X_1=-8$ otherwise,
while $X_2,\ldots,X_{10}$ are independent standard normal.
The conditional distribution of $Y \in \{1,\ldots,10\}$ given $X=x$ is multinomial with weights $w_j(x)$ defined as 
$w_j(x) = z_j(x) / \sum_{j'=1}^{p} z_{j'}(x)$, where $z_j(x) = \exp(x^T \beta_j)$ and each $\beta_j \in \mathbb{R}^{p}$ is sampled from an independent standard normal distribution.

\begin{figure}[!htb]
  \centering
  \subfigure[]{\includegraphics[width=0.495\textwidth]{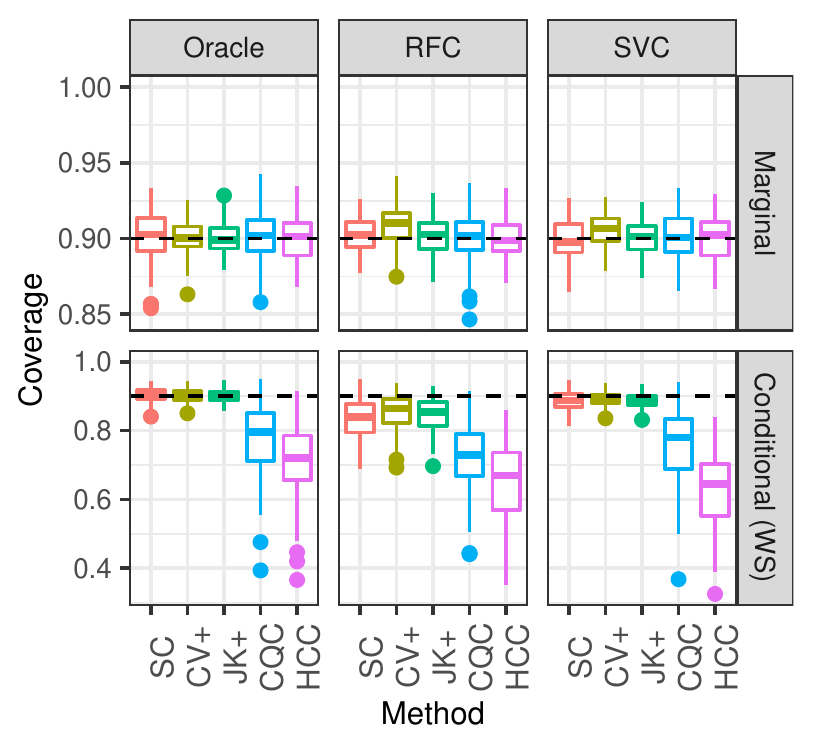}}
\subfigure[]{\includegraphics[width=0.495\textwidth]{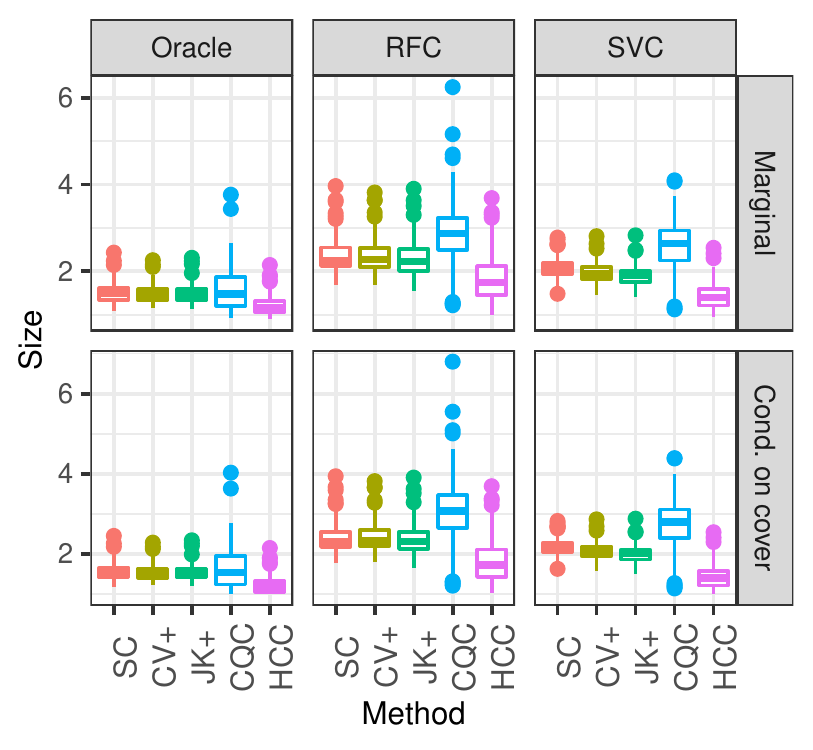}}
\caption{
Several classification methods on simulated data with 10 classes, for different choices of calibration and black-box models.
SC, CV+, and JK+ are applied with our new generalized inverse quantile conformity scores defined in \eqref{eq:define-scores}. 
The results correspond to 100 independent experiments with 1000 training samples and 5000 test samples each. All methods have 90\% marginal coverage. (a): Marginal coverage and worst-slice conditional coverage. (b): Size of prediction sets.
} 
  \label{fig:sim-1-1000}
\end{figure}

Figure~\ref{fig:sim-1-1000} confirms that our methods have valid conditional coverage if the true class probabilities are provided by an oracle.
If the probabilities are estimated by the RFC, the conditional coverage appears to be only slightly below $1-\alpha$,
and is near perfect with the SVC black box. 
By contrast, the conditional coverage of the alternative methods is always significantly lower than $1-\alpha$, even with the help of the oracle.
Our methods produce slightly larger prediction sets when the oracle is available, but our sets are typically smaller than those of CQC and only slightly larger than those of HCC when the class probabilities are estimated.
Finally, note that JK+ is the most powerful of our methods, followed by CV+, although SC is computationally more affordable.

\section{Experiments with real data}
\label{sec:real_data}

In this section, we compare the performance of our proposed methods (SC, CV+, and JK+) with the new generalized inverse quantile conformity scores defined in~\eqref{eq:define-scores} to those of HCC and CQC~\cite{cauchois2020knowing}. We found that the original suggestion of~\cite{cauchois2020knowing} to fit a quantile neural network~\cite{taylor2000quantile} on the class probability score can be unstable and yield very wide predictions. Therefore, we offer a second variant of this calibration method, denoted by CQC-RF, which replaces the quantile neural network estimator with quantile random forests~\cite{meinshausen2006quantile}; see Supplementary Section~\ref*{sec:supp-real-data} for details.

The validity and statistical efficiency of each method is evaluated according to the same metrics as in Section~\ref{sec:simulated}. In all experiments, we set $\alpha=0.1$ and use the following base predictive models: (i) kernel SVC, (ii) random forest classifier (RFC), and (iii) two-layer neural network classifier (NNet). A detailed description of each algorithm and corresponding hyper-parameters is in Supplementary Section~\ref*{sec:supp-real-data}. The methods are tested on two well-known data sets: the Mice Protein Expression data set\footnote{\url{https://archive.ics.uci.edu/ml/datasets/Mice+Protein+Expression}} and the MNIST handwritten digit data set. The supplementary material describes the processing pipeline and discusses additional experiments on the Fashion-MNIST and CIFAR10 data sets. 
Supplementary Tables~\ref*{tab:mice}--\ref*{tab:cifar10} summarize the results of our experiments in more detail and also consider additional settings.

\begin{figure}[!htb]
  \centering
  {\includegraphics[width=\textwidth]{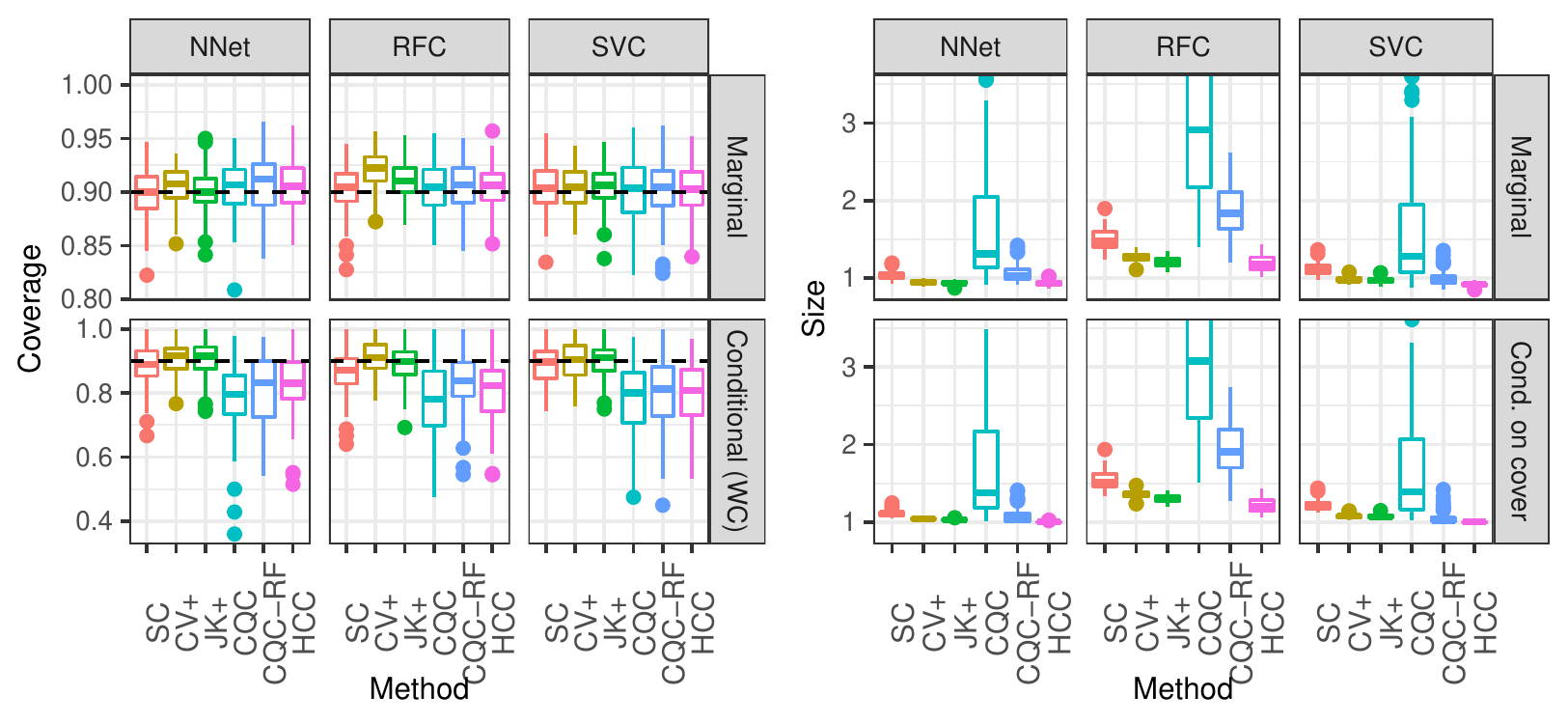}} 
\caption{Experiments on Mice Protein Expression data. 100 independent experiments with 500 randomly chosen training samples and 580 test samples each. Left: coverage. Right: size of prediction sets (extremely large values for CQC not shown).
Other details are as in Figure~\ref{fig:sim-1-1000}.
}
  \label{fig:small_sample}
\end{figure}

Figure~\ref{fig:small_sample} shows that all methods attain valid marginal coverage on the Mice Protein Expression data, as expected. Here, HCC, CQC, and CQC-RF fail to achieve conditional coverage, in contrast to the proposed methods (SC, CV+, JK+) based on our new conformity scores in~\eqref{eq:define-scores}. 
Turning to efficiency, we observe that the prediction sets of CV+ and JK+ are smaller than those of SC, and comparable in size to those of HCC. Here, the original CQC algorithm performs poorly both in terms of conditional coverage and cardinality. The CQC-RF variant is not as unstable as the original CQC, although it does not perform much better than HCC.

\begin{figure}[!htb]
  \centering
  \includegraphics[width=\textwidth]{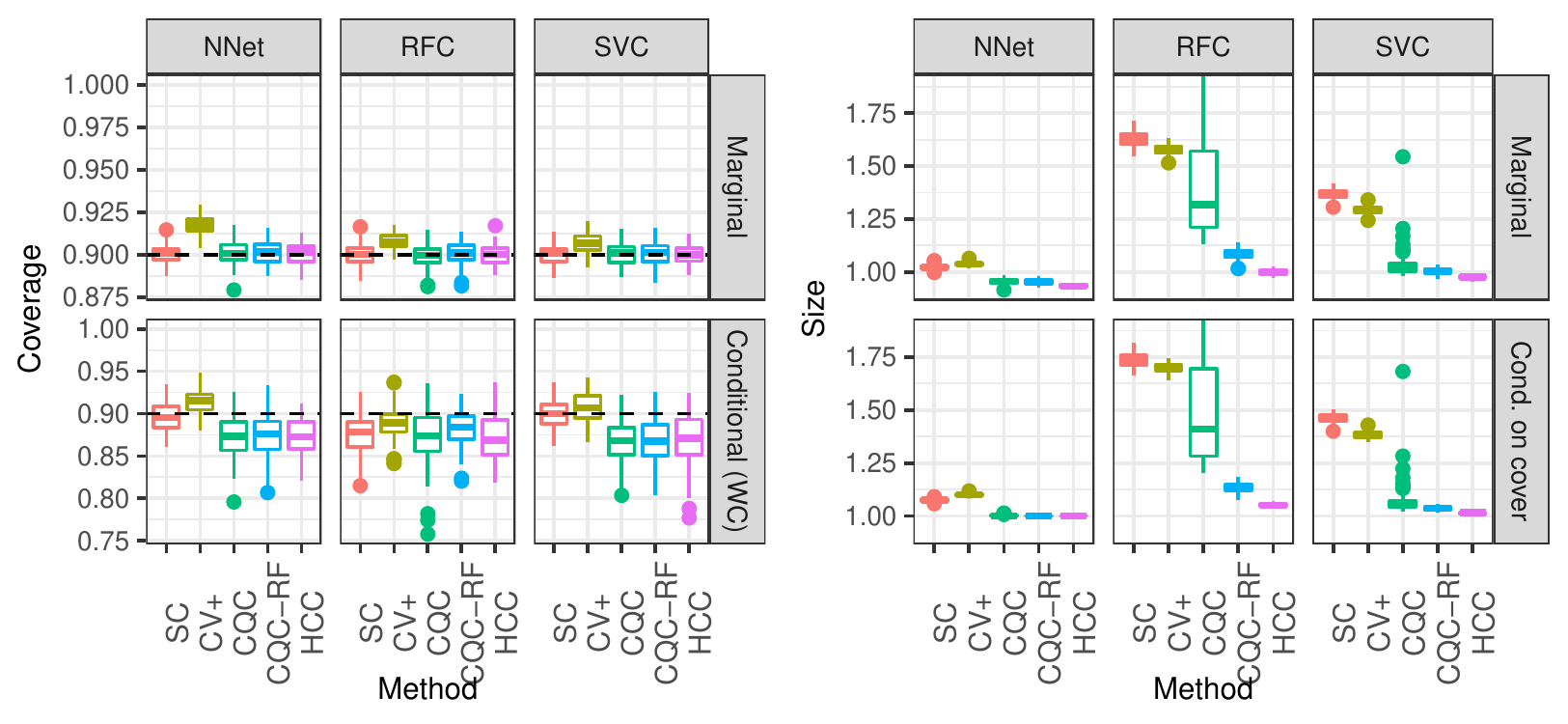}
  \includegraphics[width=\textwidth]{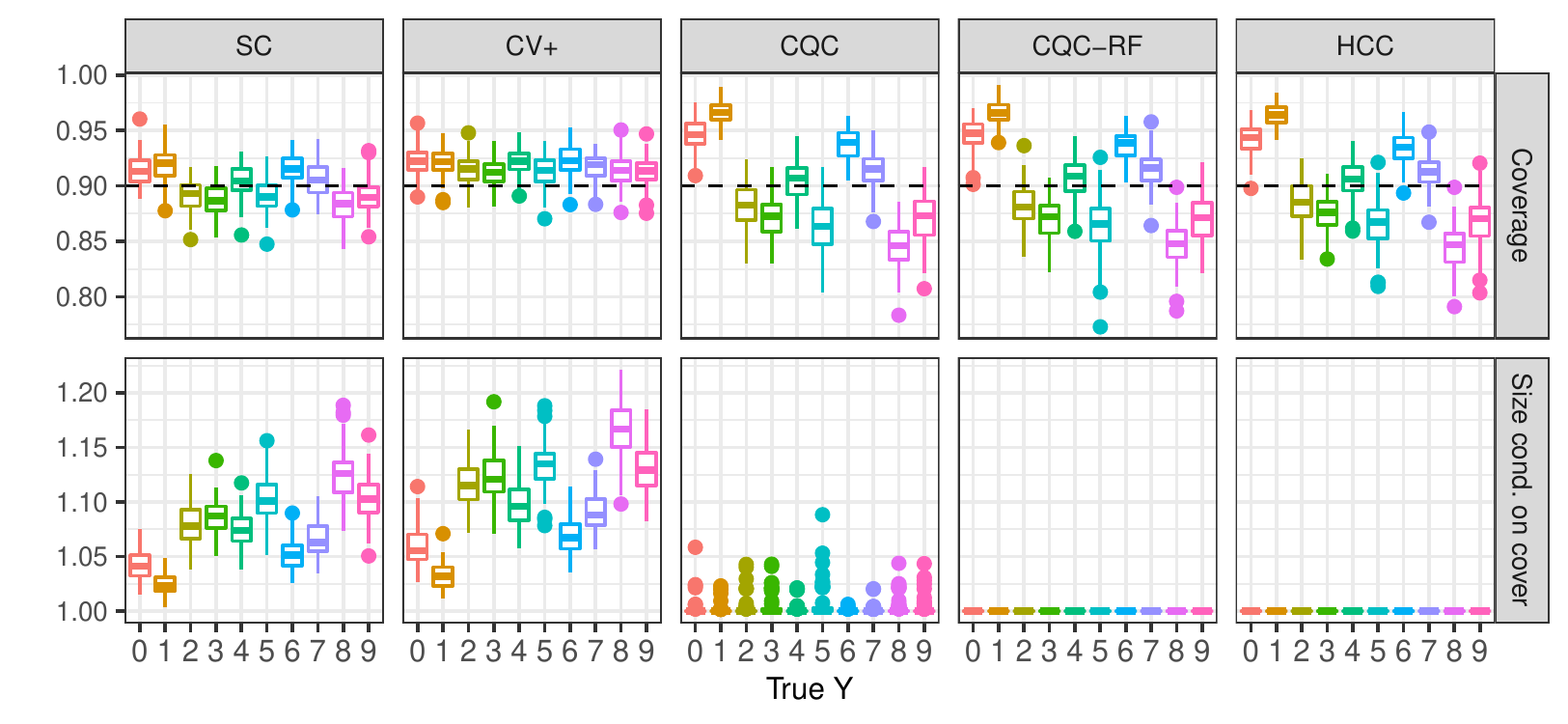}
\caption{Experiments on MNIST data. 100 independent experiments with 10000 randomly chosen training samples and 5000 test samples each. Top: coverage and size of prediction sets (large values for CQC not shown). Bottom: coverage with neural network black box, conditional on the true $Y$, and size of the corresponding prediction sets, conditional on coverage.
Other details are as in Figure~\ref{fig:sim-1-1000}.
}
  \label{fig:exp-MNIST}
\end{figure}

Figure~\ref{fig:exp-MNIST} presents the results on the MNIST data. Here, the sample size is relatively large and hence we exclude JK+ due to its higher computational cost. As in the previous experiments, all methods achieve 90\% marginal coverage. Unlike CQC, CQC-RF, and HCC, our methods also attain valid conditional coverage when relying on the NNet or SVC as base predictive models. With the RFC, all methods tend to undercover, suggesting that this classifier estimates the class probabilities poorly, and our prediction sets are larger than those constructed by CQC-RF and HCC. 
By contrast, the NNet enables our methods to achieve conditional coverage with prediction sets comparable in size to those produced by CQC-RF and HCC.
The bottom part of Figure~\ref{fig:exp-MNIST} demonstrates that CV+ also has conditional coverage given the true class label $Y$, while SC performs only slightly worse. In striking contrast, both HCC, CQC, and CQC-RF fail to achieve $90\%$ conditional coverage.

\section{Conclusions}

This paper introduced a principled and versatile modular method for constructing prediction sets for multi-class classification problems that enjoy provable finite-sample coverage, and also behave well in terms of conditional coverage when compared to alternatives. Our approach leverages the power of any black-box machine learning classifier that may be available to practitioners, and is easily calibrated via  various hold-out procedures; e.g., conformal splitting, CV+, or the jackknife+. This flexibility makes our approach widely applicable and offers options to balance between computational efficiency, data parsimony, and power.

Although this paper focused on classification, using conformity scores similar to those in~\eqref{eq:define-scores} to calibrate hold-out procedures for regression problems~\cite{romano2019conformalized,barber2019predictive} is tantalizing.
In fact, previous work in the regression setting focused on conformity scores that measure the distance of a data point from its predicted interval on the scale of the $Y$ values (which makes sense for homoscedastic regression, but may not be optimal otherwise), rather than by the amount one would need to relax the nominal threshold (our~$\tau$) until the true value is covered. We leave it to future work to explore the performance of our intuitive metrics in other settings.

The Python package at \url{https://github.com/msesia/arc} 
implements our methods.
This repository also contains code to reproduce our experiments.

\section*{Broader Impact}

Machine learning algorithms are increasingly relied upon by decision makers. It is therefore crucial to combine the predictive performance of such complex machinery with practical guarantees on the reliability and uncertainty of their output. We view the calibration methods presented in this paper as an important step towards this goal. In fact, uncertainty estimation is an effective way to quantify and communicate the benefits and limitations of machine learning. 
Moreover, the proposed methodologies provide an attractive way to move beyond the standard prediction accuracy measure used to compare algorithms. For instance, one can compare the performance of two candidate predictors, e.g., random forest and neural network (see Figure~\ref{fig:exp-MNIST}), by looking at the size of the corresponding prediction sets and/or their their conditional coverage.
Finally, the approximate conditional coverage that we seek in this work is highly relevant within the broader framework of fairness, as discussed by~\cite{Romano2020With} within a regression setting. 
While our approximate conditional coverage already implicitly reduces the risk of unwanted bias, an equalized coverage requirement~\cite{Romano2020With} can also be easily incorporated into our methods to explicitly avoid discrimination based on protected categories.

We conclude by emphasizing that the validity of our methods relies on the exchangeability of the data points. If this assumption is violated (e.g., with time-series data), our prediction sets may not have the right coverage. A general suggestion here is to always try to leverage specific knowledge of the data and of the application domain to judge whether the exchangeability assumption is reasonable. 
Finally, our data-splitting techniques in Section~\ref{sec:real_data} offer a practical way to verify empirically the validity of the predictions on any given data set.

\begin{ack}

E.~C. was partially supported by Office of Naval Research grant N00014-20-12157, and by the Army Research Office (ARO) under
grant W911NF-17-1-0304. Y.~R.~was partially supported by
ARO under the same grant.  Y.~R.~thanks the Zuckerman
Institute, ISEF Foundation, the Viterbi Fellowship, Technion, and the Koret Foundation, for providing additional research support. M.~S.~was suported by NSF grant DMS 1712800.

\end{ack}

\bibliographystyle{abbrv}
\bibliography{biblio}

\clearpage
\invisiblesection{Supplement}


\section*{Supplementary material}

# Supplementary Material for Classification with Valid and Adaptive Coverage

**Yaniv Romano**[*]
Department of Statistics
Stanford University

**Matteo Sesia**[*]
Department of Statistics
Stanford University

**Emmanuel J. Candès**
Departments of Mathematics
and of Statistics
Stanford University

## S1 Supplementary methods

### S1.1 Adaptive classification with minimax jackknife+ calibration

We can apply the minimax calibration technique of [2] to obtain a non-trivial variation of Algorithm 2 for which marginal coverage can be rigorously proved at level $1 - \alpha$, without modifying the current input level. Here, we consider the jackknife+—i.e., $K = n$—for simplicity. The only difference with Algorithm 2 is that the prediction set in (11) is replaced by the following larger set:

$$\hat{\mathcal{C}}_{n,\alpha}^{\text{J+mm}}(X_{n+1}) = \left\{ y \in \mathcal{Y} : \right. \\ \left. \sum_{i=1}^{n} \mathbf{1}\left[ E(X_i, Y_i, U_i; \hat{\pi}^i) < \min_{j \in \{1,...,n\}} E(X_{n+1}, y, U_{n+1}; \hat{\pi}^j) \right] < (1-\alpha)(n+1) \right\}. \quad \text{(S1)}$$

**Theorem S1.** *Under the same assumptions of Theorem 1, the output of Algorithm 2 with $K = n$, and* (11) *replaced by* (S1), *satisfies:*

$$\mathbb{P}\left[ Y_{n+1} \in \hat{\mathcal{C}}_{n,\alpha}^{\text{J+mm}}(X_{n+1}) \right] \geq 1 - \alpha. \quad \text{(S2)}$$

### S1.2 Quantifying conditional coverage in finite samples

Similarly to the approach of [1], we measure coverage over a slab

$$S_{v,a,b} = \{ x \in \mathbb{R}^p : a \leq v^T x \leq b \}$$

of the feature space, where the values of $v \in \mathbb{R}^p$ and $a < b \in \mathbb{R}$ are chosen adversarially but independently of the data. In particular, for any fixed classification prediction set $\hat{\mathcal{C}}$ and $\delta \in (0,1)$, we define

$$\text{WSC}(\hat{\mathcal{C}}; \delta) = \inf_{v \in \mathbb{R}^p,\ a<b \in \mathbb{R}} \left\{ \mathbb{P}[Y \in \hat{\mathcal{C}}(X) \mid X \in S_{v,a,b}] \text{ s.t. } \mathbb{P}[X \in S_{v,a,b}] \geq \delta ] \right\}.$$

In practice, we estimate WSC for a particular $\hat{\mathcal{C}}$ by sampling 1000 independent vectors $v$ on the unit sphere in $\mathbb{R}^p$ and optimizing the corresponding parameters $a, b$ through a grid search; we set $\delta = 0.1$. To avoid finite-sample negative bias, we partition the test data into two subsets (e.g., containing 25% and 75% of the samples respectively); then, we use the first subset to estimate the optimal values $v^*, a^*, b^*$, and the second subset to evaluate conditional coverage:

$$\mathbb{P}[Y \in \hat{\mathcal{C}}(X) \mid X \in S_{v^*, a^*, b^*}]. \quad \text{(S3)}$$

---

[*]Equal contribution.

Therefore, regardless of the quality of our solution $v^*, a^*, b^*$ to the above optimization problem, the quantity in (S3) should be equal to the nominal coverage level $1 - \alpha$ for any method with valid conditional coverage. However, it is worth highlighting that controlling (S3) does not necessarily imply that conditional coverage holds more generally, which is why we also look at alternative measures of conditional coverage given either the value of certain features (e.g., $X_1$), or that of the true label $Y$.

## S2 Supplementary proofs

*Proof of Theorem 1.* We begin by proving the lower bound on coverage. By construction of the prediction set in (8), we know that

$$Y_{n+1} \in \hat{\mathcal{C}}_{n,\alpha}^{\text{sc}}(X_{n+1})$$

if and only if

$$\min\{\tau \in [0,1] : Y_{n+1} \in \mathcal{S}(X_{n+1}, U_{n+1}; \hat{\pi}, \tau)\} \leq \hat{Q}_{1-\alpha}(\{E_i\}_{i \in \mathcal{I}_2}),$$

or, equivalently, if and only if

$$E_{n+1} \leq \hat{Q}_{1-\alpha}(\{E_i\}_{i \in \mathcal{I}_2}). \tag{S4}$$

Since all the conformity scores $E_{n+1}$ and $\{E_i\}_{i \in \mathcal{I}_2}$ are exchangeable, the probability of the event in (S4) can be no larger than $1 - \alpha$. The formal proof of this statement is standard at this point, so we simply refer to [3] for the remaining technical details. The proof for the upper bound also immediately follows from (S4) by applying Lemma 2 in [3]. □

*Proof of Theorem 2.* The proof is essentially an application of the main result in [2]. This will become apparent after we reduce our claim to the setting in the aforementioned paper. We now examine this reduction.

Imagine that we have access to $m = n/K$ test points

$$(X_{n+1}, Y_{n+1}, U_{n+1}), \ldots, (X_{n+m}, Y_{n+m}, U_{n+m})$$

as well as the training data; we will call this data set the *augmented* data set. After partitioning the training data into sets $\mathcal{I}_1, \ldots, \mathcal{I}_K$ of size $m$, we define $\mathcal{I}_{K+1} = \{n+1, \ldots, n+m\}$ as the set of test points. For any distinct $k, k' \in \{1, \ldots, K+1\}$, let $\tilde{\pi}^{k,k'}$ define the class probability estimator obtained by fitting the black box on the data in $\{1, \ldots, n+m\} \setminus (\mathcal{I}_k \cup \mathcal{I}_{k'})$. Note that $\tilde{\pi}^{k,K+1} = \hat{\pi}^k$ for any $k$.

Next, define the matrix $R \in \mathbb{R}^{(n+m) \times (n+m)}$ with entries

$$R_{i,j} = \begin{cases} +\infty, & \text{if } k(i) = k(j), \\ E(X_i, Y_i, U_i; \tilde{\pi}^{k(i), k(j)}), & \text{if } k(i) \neq k(j), \end{cases}$$

and the comparison matrix $A \in \{0,1\}^{(n+m) \times (n+m)}$ with entries

$$A_{ij} = \mathbf{1}[R_{ij} > R_{ji}]. \tag{S5}$$

Note that

$$Y_{n+1} \notin \hat{\mathcal{C}}_{n,\alpha}^{\text{CV+}}(X_{n+1}) \iff (n+1) \in \mathcal{F}(A),$$

where the set $\mathcal{F}(A)$ is defined as in [2]:

$$\mathcal{F}(A) = \left\{ i \in \{1, \ldots, n+m\} : \sum_{j=1}^{n+m} A_{i,j} \geq (1-\alpha)(n+1) \right\}. \tag{S6}$$

The rest of the proof follows directly by applying Lemma S1 below, which is established by the proof of Theorem 4 in [2]. To invoke this lemma, we only need to check that $A \overset{d}{=} \Sigma A \Sigma^\top$, where $A$ is defined as in (S5), and $\Sigma$ is any permutation matrix that does not mix points assigned to different

folds (so that the unordered set of probability estimators $\{\hat{\pi}^k\}_{k=1}^{K}$ is invariant). This is easy to verify. Let $\sigma(1), \ldots, \sigma(n+m)$ be the permutation of the data points corresponding to $\Sigma$, so that

$$(\Sigma A \Sigma^\top)_{ij} = A_{\sigma(i)\sigma(j)}.$$

Then, for any $i, j$ such that $k(i) \neq k(j)$,

$$\begin{aligned}
A_{\sigma(i)\sigma(j)} &= \mathbf{1}\left[E(X_{\sigma(i)}, Y_{\sigma(i)}, U_{\sigma(i)}; \tilde{\pi}^{k(\sigma(i)),k(\sigma(j))}) > E(X_{\sigma(j)}, Y_{\sigma(j)}, U_{\sigma(j)}; \tilde{\pi}^{k(\sigma(i)),k(\sigma(j))})\right] \\
&= \mathbf{1}\left[E(X_{\sigma(i)}, Y_{\sigma(i)}, U_{\sigma(i)}; \tilde{\pi}^{k(i),k(j)}) > E(X_{\sigma(j)}, Y_{\sigma(j)}, U_{\sigma(j)}; \tilde{\pi}^{k(i),k(j)})\right] \\
&\stackrel{d}{=} \mathbf{1}\left[E(X_i, Y_i, U_i; \tilde{\pi}^{k(i),k(j)}) > E(X_j, Y_j, U_j; \tilde{\pi}^{k(i),k(j)})\right] \\
&= A_{ij}.
\end{aligned}$$

Above, the second equality holds because the black-box estimators $\hat{\pi}^k$ are invariant to the ordering of their input data points, and the third equality in distribution holds because the data points $(X_i, Y_i, U_i)$ are exchangeable. Finally, we also trivially know that $A_{\sigma(i)\sigma(j)} = A_{ij}$ for any $i, j$ such that $k(i) = k(j)$.

□

**Lemma S1** (Proved in [2]). *Consider any partition of $\{1, \ldots, n+m\}$ points into $K+1$ folds $\mathcal{I}_1, \ldots, \mathcal{I}_{K+1}$, with $m = n/K$. If a random matrix $A \in \{0,1\}^{(n+m) \times (n+m)}$ satisfies $A \stackrel{d}{=} \Sigma A \Sigma^\top$ for any $(n+m) \times (n+m)$ permutation matrix $\Sigma$ that does not mix points assigned to different folds, then, for any fixed $\alpha \in (0,1)$,*

$$\mathbb{P}\left[(n+1) \in \mathcal{F}(A)\right] \leq 2\alpha + \min\left\{\frac{2(1-1/K)}{n/K+1}, \frac{1-K/n}{K+1}\right\}, \tag{S7}$$

*where the set $\mathcal{F}(A)$ is defined as in* (S6) *and depends on $\alpha$. In the special case where $K = n$, this bound simplifies to:*

$$\mathbb{P}\left[(n+1) \in \mathcal{F}(A)\right] \leq 2\alpha. \tag{S8}$$

*Proof of Theorem S1.* The proof is effectively identical to that of Theorem 3 in [2], by the same argument as in the proof of Theorem 2.

□

## S3 Supplementary experiments with simulated data

### S3.1 Implementation details for black-box classifiers

We have applied the following black-box classification methods to estimate label probabilities:

- a support vector classifier (SVC) with linear kernel, as implemented by the `sklearn` Python package with default parameters;
- a random forest classifier (RFC) with 1000 estimators of maximum depth 5, as implemented by the `sklearn` Python package with default parameters (except for the maximum number of features considered at each split, which we set equal to $p$).

For the CQC method, we carry out quantile regression on the classification scores using the same deep neural network employed in [3].

### S3.2 Experiments with multinomial model and inhomogeneous features

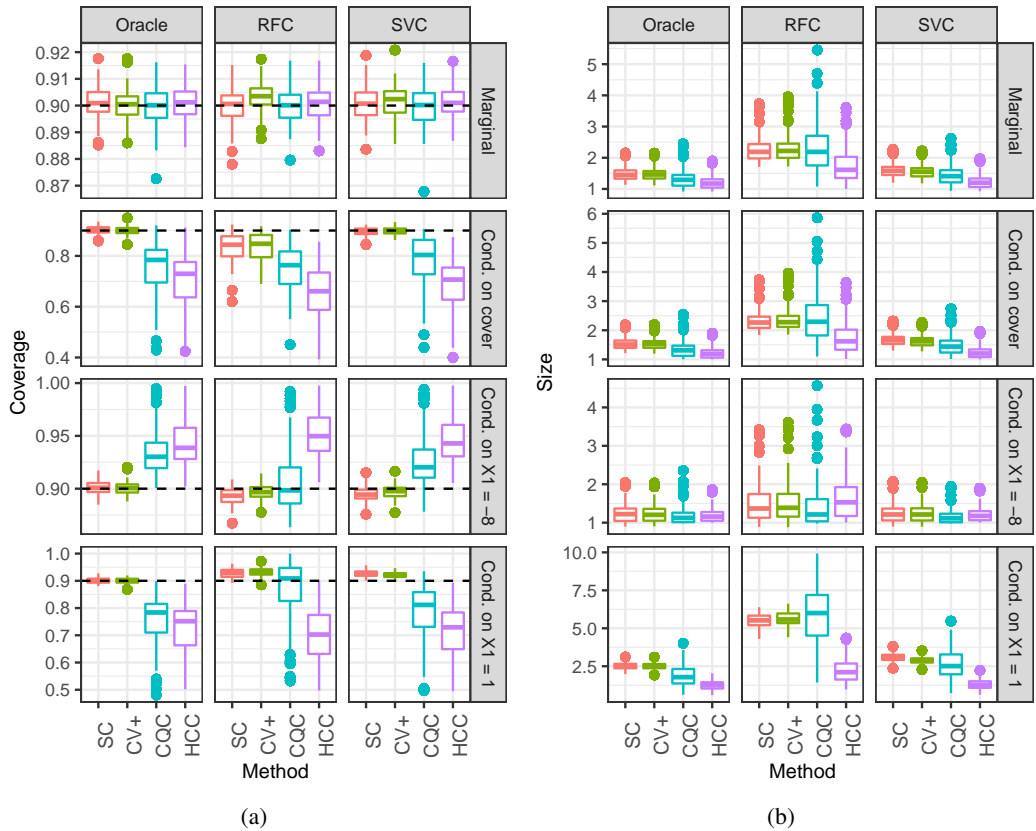

Figure S1: Different classification methods on simulated data with 10 classes, for different choices of calibration and black-box models. The results correspond to 100 independent experiments with 10000 training samples and 5000 test samples each. JK+ is omitted for computational reasons. Other details are as in Figure 1.

### S3.3 Experiments with heteroscedastic decision-tree model and discrete features

We set $p = 5$ and generate each sample of features $X \in \mathbb{R}^p$ independently as follows: $X_1 = +1$ w.p. $3/4$, and $X_1 = -1$ w.p. $1/4$; $X_2 = +1$ w.p. $3/4$, and $X_2 = -2$ w.p. $1/4$; $X_3 = +1$ w.p. $1/4$, and $X_3 = -2$ w.p. $1/2$; $X_4$ is uniformly distributed on $\{1, \ldots, 4\}$; and $X_5 \sim \mathcal{N}(0, 1)$. The labels $Y$ belong to one of 4 possible classes, and their conditional distribution given $X = x$ is given by the decision tree shown in Figure S2, which only depends on the first four features.

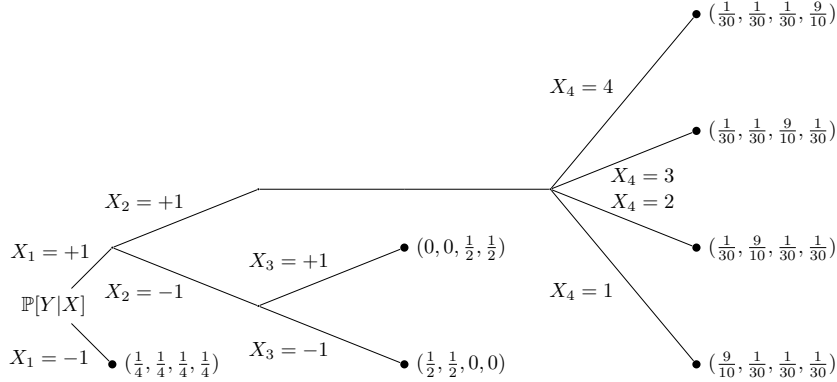

Figure S2: A toy model for $P_{Y|X}$ in a classification setting with 4 labels.

The performances of the different methods on data generated from this model are compared in Figure S3. Here, the size of the training sample is equal to 10000 and the size of the test sample is equal to 5000; all experiments are repeated 100 times. Since these training sets are fairly large, for computational convenience we do not apply the JK+ method; see Figure S4 for a comparison including JK+ with smaller sample sizes.

These results are qualitatively consistent with those from Section 3.2, confirming that our methods have good approximate conditional coverage compared to the alternatives while not suffering from a significant power loss. It is interesting to note that the conditional distribution of $Y \mid X$ is more complicated here than in the previous example, hence the reason for a larger sample size. Despite this large sample size, the SVC black-box is unable to learn good estimates of the class probabilities. This is why methods with marginal coverage have relatively low power and poor conditional coverage. By contrast, the RFC black-box can learn these class probabilities quite accurately, and thus it allows our SC and CV+ methods to perform on par with the oracle (especially CV+, as expected). Again, the alternative methods do not achieve conditional coverage even with the help of the oracle.

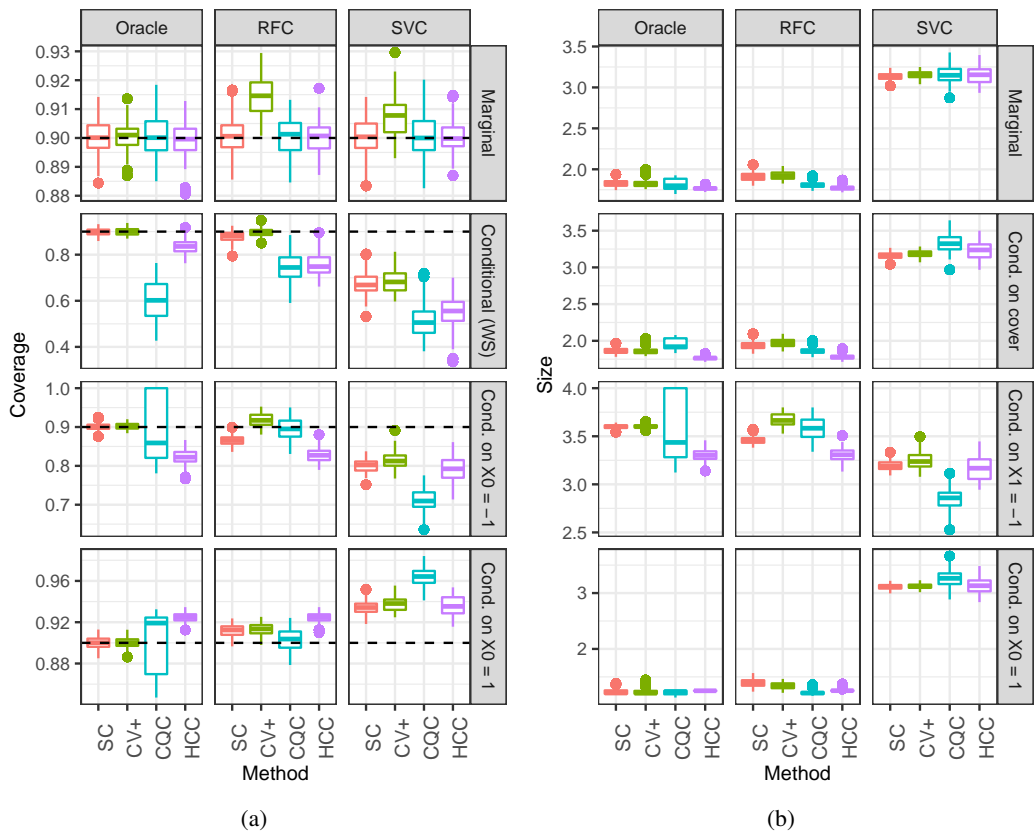

Figure S3: Performance of alternative classification methods on simulated data with 4 classes. Results from 100 independent experiments with 10000 training samples and 5000 test samples each. Other details as in Figure 1.

6

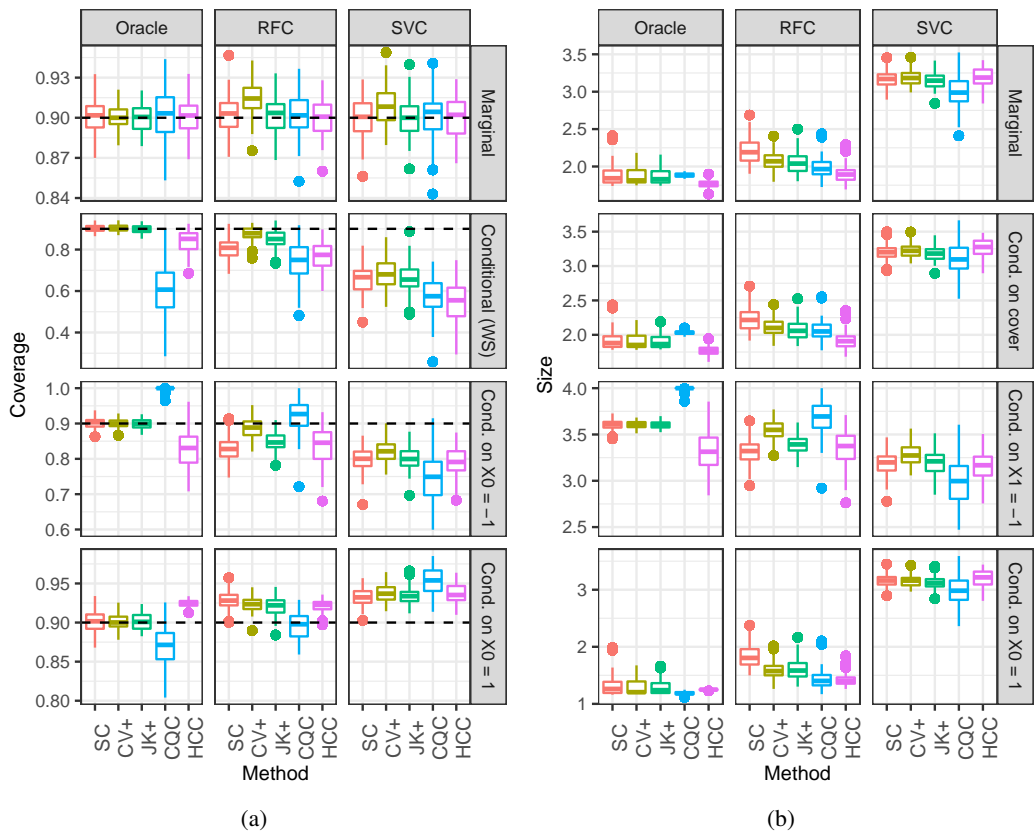

Figure S4: Performance of alternative classification methods on simulated data with 4 classes. Results from 100 independent experiments with 1000 training samples and 5000 test samples each. Other details are as in Figure S3.

7

## S4 Supplementary experiments with real data

We compare the performance of our methods to that of HCC and CQC on four popular benchmark data sets:

1. MNIST is a handwritten digit classification data set, containing 60000 grayscale images of size $28 \times 28$ pixels, each associated with one of $C = 10$ classes. As a pre-processing step, we apply Principal Component Analysis (PCA) to each image, resulting in a feature vector $X$ of length $p = 50$.

2. CIFAR10 is another image classification data set. The data includes 50000 RGB images, each of size $32 \times 32 \times 3$, belonging to one of $C = 10$ classes. We also use PCA to reduce the dimension to $p = 50$.

3. Fashion-MNIST contains 60000 images associated with $C = 10$ classes of clothes. We run the same pre-processing step as in MNIST and CIFAR10, resulting in $p = 50$ features.

4. The task in the Mice Protein Expression data set[2] is to identify the class of a mouse based on genetic, behavioral and treatment covariates. After applying standard data cleaning, we have 1080 samples, $p = 77$ features, and $C = 8$ classes.

We use the same baseline predictive algorithms as in Section S3.1, although with slightly different RFC parameters—here the number of estimators is 100, and the minimum number of samples at a leaf node is 3. Additionally, we also consider a neural network (NNet) with one hidden layer of size 64 and ReLU activation function. We use the adam optimizer, with a minibatch of size 128, a learning rate of 0.01, and a total number of epochs equal to 20. The CQC method is implemented as described in Section S3.1. In the real data experiments we present a second variant of CQC, namely CQC-RF, where we replace the quantile neural network algorithm with quantile random forest. To this end, we use the default `skgarden` hyper-parameters for quantile random forest, except for the number of estimators and the minimum number of samples required to split an internal node, which we set to 100 and 3, respectively.

In the numerical experiments, we set the target coverage level to $90\%$ and compare the coverage, conditional coverage, and length of the different calibration methods combined with the above predictive algorithms. The performance metrics are averaged over 100 experiments. The results in Table S1 are obtained by randomly selecting $n_{\text{train}} \in \{500, 1000\}$ training examples from the Mice Protein Expression data set, used to fit and calibrate the predictive models. The remaining $n_{\text{test}} \in \{580, 80\}$ unseen samples formulate a test set, in which we evaluate the methods' performance. Tables S2, S3, and S4 correspond to MNIST, Fashion-MNIST, and CIFAR10 data sets. Each experiment is conducted by randomly selecting $n_{\text{train}} \in \{500, 1000, 5000, 10000\}$ training examples as well as a disjoint set of 5000 unseen test samples, selected at random.

In sum, all the calibration methods achieve an exact $90\%$ marginal coverage, as guaranteed by the theory. CV+ and JK+ tend to achieve conditional coverage as well (green colored numbers), and SC performs slightly worse. In contrast, in most cases CQC, CQC-RF, and HCC fail (red colored numbers) to obtain the desired conditional coverage. As for the statistical efficiently, HCC often results in the shortest prediction sets—while failing to attain conditional coverage. Here, our methods are typically competitive and can even produce smaller prediction sets in some cases.

## S5 Supplementary tables

---

[2] https://archive.ics.uci.edu/ml/datasets/Mice+Protein+Expression

Table S1: Results of experiments on the Mice Protein Expression data. Mean marginal coverage, conditional coverage (worst-slab), and average size of the prediction sets conditional on coverage. Experiments obtained with different training set sizes are shown in different groups of rows. Standard deviations are shown in round brackets. Red: conditional coverage values below 0.88; green: conditional coverage values above 0.9; blue: smallest mean prediction set conditional on coverage.

| | | Coverage | | | | | | Conditional coverage | | | | | | Size \| coverage | | | | | |
|---|---|---|---|---|---|---|---|---|---|---|---|---|---|---|---|---|---|---|---|
| Black box | SC | CV+ | JK+ | CQC | CQC-RF | HCC | SC | CV+ | JK+ | CQC | CQC-RF | HCC | SC | CV+ | JK+ | CQC | CQC-RF | HCC |
| **500** | | | | | | | | | | | | | | | | | | |
| NNet | 0.90 (0.02) | 0.90 (0.02) | 0.90 (0.02) | 0.90 (0.02) | 0.91 (0.03) | 0.91 (0.02) | 0.89 (0.06) | 0.91 (0.05) | 0.91 (0.06) | 0.79 (0.11) | 0.81 (0.10) | 0.83 (0.09) | 1.11 (0.03) | 1.04 (0.01) | 1.03 (0.01) | 1.96 (1.23) | 1.08 (0.10) | 1.00 (0.00) |
| RFC | 0.90 (0.02) | 0.92 (0.02) | 0.91 (0.02) | 0.90 (0.02) | 0.91 (0.02) | 0.90 (0.02) | 0.86 (0.07) | 0.91 (0.05) | 0.89 (0.06) | 0.78 (0.12) | 0.83 (0.09) | 0.80 (0.10) | 1.54 (0.12) | 1.36 (0.05) | 1.30 (0.05) | 3.43 (1.33) | 1.95 (0.33) | 1.22 (0.09) |
| SVC | 0.90 (0.02) | 0.90 (0.02) | 0.90 (0.02) | 0.90 (0.03) | 0.90 (0.03) | 0.90 (0.02) | 0.89 (0.06) | 0.90 (0.06) | 0.90 (0.05) | 0.78 (0.12) | 0.80 (0.11) | 0.80 (0.10) | 1.22 (0.06) | 1.08 (0.02) | 1.07 (0.02) | 1.93 (1.24) | 1.05 (0.07) | 1.00 (0.00) |
| **1000** | | | | | | | | | | | | | | | | | | |
| NNet | 0.89 (0.02) | 0.91 (0.02) | 0.91 (0.04) | 0.90 (0.04) | 0.90 (0.03) | 0.90 (0.02) | NA (NA) | 0.91 (0.13) | NA (NA) | 0.91 (0.11) | NA (NA) | NA (NA) | 1.03 (0.02) | 1.01 (0.01) | 1.01 (0.01) | 1.00 (0.00) | 1.00 (0.00) | 1.00 (0.00) |
| RFC | 0.90 (0.03) | 0.90 (0.03) | 0.89 (0.03) | 0.89 (0.04) | 0.89 (0.04) | 0.90 (0.03) | NA (NA) | NA (NA) | NA (NA) | 0.92 (0.13) | NA (NA) | NA (NA) | 1.31 (0.06) | 1.19 (0.04) | 1.17 (0.06) | 1.92 (0.79) | 1.19 (0.10) | 1.01 (0.01) |
| SVC | 0.89 (0.04) | 0.90 (0.03) | 0.90 (0.03) | 0.90 (0.03) | 0.91 (0.04) | 0.90 (0.03) | NA (NA) | NA (NA) | NA (NA) | NA (NA) | NA (NA) | NA (NA) | 1.06 (0.04) | 1.02 (0.02) | 1.02 (0.02) | 1.10 (0.65) | 1.00 (0.00) | 1.00 (0.00) |

9

Table S2: Results of experiments on MNIST data. Other details are as in Table S1.

| Black box | Coverage | | | | | | | Conditional coverage | | | | | | | Size \| coverage | | | | | | |
|---|---|---|---|---|---|---|---|---|---|---|---|---|---|---|---|---|---|---|---|---|---|
| | SC | CV+ | JK+ | CQC | CQC-RF | HCC | | SC | CV+ | JK+ | CQC | CQC-RF | HCC | | SC | CV+ | JK+ | CQC | CQC-RF | HCC | |
| **500** | | | | | | | | | | | | | | | | | | | | | |
| NNet | 0.90 (0.02) | 0.92 (0.01) | 0.90 (0.01) | 0.90 (0.02) | 0.90 (0.02) | 0.90 (0.02) | | 0.85 (0.04) | 0.90 (0.03) | 0.88 (0.03) | 0.81 (0.05) | 0.86 (0.04) | 0.86 (0.04) | | 2.51 (0.68) | 1.67 (0.08) | 1.54 (0.07) | 6.63 (1.29) | 3.17 (0.61) | 2.25 (0.39) | |
| RFC | 0.91 (0.02) | 0.93 (0.01) | 0.92 (0.01) | 0.90 (0.02) | 0.90 (0.02) | 0.90 (0.02) | | 0.88 (0.03) | 0.88 (0.03) | 0.88 (0.03) | 0.80 (0.07) | 0.85 (0.04) | 0.86 (0.03) | | 2.55 (0.33) | 2.21 (0.13) | 2.15 (0.13) | 7.14 (1.77) | 4.02 (0.89) | 2.06 (0.31) | |
| SVC | 0.90 (0.02) | 0.92 (0.01) | 0.90 (0.01) | 0.90 (0.02) | 0.90 (0.02) | 0.90 (0.02) | | 0.87 (0.04) | 0.91 (0.02) | 0.89 (0.02) | 0.79 (0.05) | 0.83 (0.05) | 0.85 (0.04) | | 2.51 (0.29) | 2.19 (0.12) | 2.06 (0.13) | 6.68 (1.52) | 3.97 (0.89) | 1.91 (0.31) | |
| **1000** | | | | | | | | | | | | | | | | | | | | | |
| NNet | 0.90 (0.01) | 0.93 (0.01) | 0.91 (0.01) | 0.90 (0.02) | 0.90 (0.02) | 0.90 (0.01) | | 0.88 (0.03) | 0.92 (0.02) | 0.90 (0.02) | 0.83 (0.05) | 0.86 (0.04) | 0.86 (0.03) | | 1.53 (0.09) | 1.43 (0.04) | 1.29 (0.03) | 5.37 (1.47) | 1.74 (0.24) | 1.36 (0.11) | |
| RFC | 0.90 (0.01) | 0.92 (0.01) | 0.92 (0.01) | 0.90 (0.02) | 0.90 (0.02) | 0.90 (0.01) | | 0.87 (0.03) | 0.88 (0.03) | 0.88 (0.03) | 0.81 (0.05) | 0.87 (0.03) | 0.86 (0.04) | | 2.07 (0.13) | 1.99 (0.08) | 1.96 (0.08) | 6.17 (1.59) | 2.47 (0.44) | 1.53 (0.13) | |
| SVC | 0.90 (0.01) | 0.92 (0.01) | 0.90 (0.01) | 0.90 (0.02) | 0.90 (0.02) | 0.90 (0.01) | | 0.89 (0.03) | 0.92 (0.02) | 0.90 (0.02) | 0.81 (0.04) | 0.86 (0.04) | 0.86 (0.03) | | 2.07 (0.11) | 1.91 (0.08) | 1.82 (0.08) | 5.63 (1.58) | 1.89 (0.36) | 1.37 (0.11) | |
| **5000** | | | | | | | | | | | | | | | | | | | | | |
| NNet | 0.90 (0.01) | 0.92 (0.01) | NA | 0.90 (0.01) | 0.90 (0.01) | 0.90 (0.01) | | 0.90 (0.02) | 0.92 (0.01) | NA | 0.87 (0.03) | 0.87 (0.03) | 0.88 (0.03) | | 1.13 (0.01) | 1.14 (0.01) | NA | 1.14 (0.46) | 1.02 (0.02) | 1.00 (0.00) | |
| RFC | 0.90 (0.01) | 0.91 (0.00) | NA | 0.90 (0.01) | 0.90 (0.01) | 0.90 (0.01) | | 0.87 (0.03) | 0.89 (0.02) | NA | 0.85 (0.04) | 0.88 (0.02) | 0.87 (0.03) | | 1.80 (0.05) | 1.76 (0.03) | NA | 3.05 (1.19) | 1.27 (0.06) | 1.11 (0.02) | |
| SVC | 0.90 (0.01) | 0.91 (0.01) | NA | 0.90 (0.01) | 0.90 (0.01) | 0.90 (0.01) | | 0.90 (0.02) | 0.91 (0.02) | NA | 0.85 (0.04) | 0.87 (0.02) | 0.87 (0.03) | | 1.61 (0.04) | 1.50 (0.03) | NA | 1.96 (1.28) | 1.09 (0.02) | 1.05 (0.01) | |
| **10000** | | | | | | | | | | | | | | | | | | | | | |
| NNet | 0.90 (0.01) | 0.92 (0.00) | NA | 0.90 (0.01) | 0.90 (0.01) | 0.90 (0.01) | | 0.90 (0.02) | 0.92 (0.01) | NA | 0.87 (0.02) | 0.87 (0.02) | 0.87 (0.02) | | 1.07 (0.01) | 1.10 (0.01) | NA | 1.00 (0.00) | 1.00 (0.00) | 1.00 (0.00) | |
| RFC | 0.90 (0.01) | 0.91 (0.00) | NA | 0.90 (0.01) | 0.90 (0.01) | 0.90 (0.01) | | 0.88 (0.02) | 0.89 (0.02) | NA | 0.87 (0.03) | 0.88 (0.02) | 0.87 (0.03) | | 1.74 (0.03) | 1.70 (0.02) | NA | 1.59 (0.60) | 1.13 (0.02) | 1.05 (0.01) | |
| SVC | 0.90 (0.01) | 0.91 (0.01) | NA | 0.90 (0.01) | 0.90 (0.01) | 0.90 (0.01) | | 0.90 (0.02) | 0.91 (0.02) | NA | 0.87 (0.03) | 0.87 (0.03) | 0.87 (0.03) | | 1.46 (0.02) | 1.38 (0.02) | NA | 1.07 (0.07) | 1.04 (0.01) | 1.02 (0.00) | |

10

Table S3: Results of experiments on Fashion-MNIST data. Other details are as in Table S1.

| Black box | Coverage | | | | | | | Conditional coverage | | | | | | | Size / coverage | | | | | | |
|---|---|---|---|---|---|---|---|---|---|---|---|---|---|---|---|---|---|---|---|---|---|
| | SC | CV+ | JK+ | CQC | CQC-RF | HCC | | SC | CV+ | JK+ | CQC | CQC-RF | HCC | | SC | CV+ | JK+ | CQC | CQC-RF | HCC | |
| **500** | | | | | | | | | | | | | | | | | | | | | |
| NNet | 0.90 (0.02) | 0.92 (0.01) | 0.90 (0.01) | 0.91 (0.03) | 0.91 (0.02) | 0.90 (0.02) | | 0.86 (0.04) | 0.88 (0.03) | 0.86 (0.03) | 0.79 (0.06) | 0.85 (0.04) | 0.85 (0.04) | | 2.97 (0.78) | 1.96 (0.22) | 1.82 (0.17) | 6.59 (1.24) | 3.59 (0.69) | 2.60 (0.55) | |
| RFC | 0.90 (0.02) | 0.92 (0.01) | 0.92 (0.01) | 0.90 (0.03) | 0.90 (0.02) | 0.91 (0.02) | | 0.88 (0.03) | 0.91 (0.02) | 0.90 (0.02) | 0.76 (0.07) | 0.82 (0.06) | 0.86 (0.05) | | 2.47 (0.24) | 2.28 (0.11) | 2.23 (0.10) | 6.68 (1.77) | 3.95 (0.83) | 1.98 (0.22) | |
| SVC | 0.90 (0.02) | 0.93 (0.01) | 0.90 (0.02) | 0.90 (0.03) | 0.90 (0.02) | 0.90 (0.02) | | 0.86 (0.03) | 0.91 (0.02) | 0.88 (0.03) | 0.76 (0.06) | 0.80 (0.05) | 0.84 (0.05) | | 2.74 (0.29) | 2.43 (0.13) | 2.30 (0.14) | 6.61 (1.41) | 4.87 (0.80) | 2.15 (0.29) | |
| **1000** | | | | | | | | | | | | | | | | | | | | | |
| NNet | 0.90 (0.01) | 0.93 (0.01) | 0.90 (0.01) | 0.90 (0.02) | 0.90 (0.02) | 0.90 (0.01) | | 0.85 (0.04) | 0.91 (0.02) | 0.87 (0.03) | 0.81 (0.06) | 0.85 (0.04) | 0.85 (0.04) | | 1.82 (0.21) | 1.71 (0.06) | 1.53 (0.05) | 5.73 (1.57) | 2.13 (0.31) | 1.72 (0.15) | |
| RFC | 0.90 (0.01) | 0.92 (0.01) | 0.91 (0.01) | 0.90 (0.01) | 0.90 (0.02) | 0.90 (0.01) | | 0.89 (0.03) | 0.91 (0.02) | 0.90 (0.02) | 0.77 (0.06) | 0.84 (0.04) | 0.85 (0.04) | | 2.20 (0.13) | 2.11 (0.07) | 2.09 (0.07) | 5.88 (1.73) | 2.70 (0.55) | 1.64 (0.10) | |
| SVC | 0.90 (0.02) | 0.92 (0.01) | 0.90 (0.01) | 0.90 (0.02) | 0.90 (0.02) | 0.90 (0.01) | | 0.88 (0.02) | 0.91 (0.02) | 0.88 (0.02) | 0.78 (0.05) | 0.83 (0.04) | 0.85 (0.04) | | 2.30 (0.13) | 2.11 (0.07) | 2.01 (0.08) | 5.89 (1.60) | 3.26 (0.54) | 1.71 (0.12) | |
| **5000** | | | | | | | | | | | | | | | | | | | | | |
| NNet | 0.90 (0.01) | 0.94 (0.01) | NA | 0.90 (0.01) | 0.90 (0.01) | 0.90 (0.01) | | 0.86 (0.03) | 0.93 (0.02) | NA | 0.84 (0.04) | 0.86 (0.03) | 0.86 (0.03) | | 1.35 (0.02) | 1.45 (0.02) | NA | 3.29 (1.40) | 1.36 (0.05) | 1.29 (0.04) | |
| RFC | 0.90 (0.01) | 0.91 (0.01) | NA | 0.90 (0.01) | 0.90 (0.01) | 0.90 (0.01) | | 0.89 (0.02) | 0.91 (0.02) | NA | 0.81 (0.06) | 0.87 (0.03) | 0.86 (0.03) | | 1.92 (0.04) | 1.87 (0.02) | NA | 3.91 (1.50) | 1.65 (0.10) | 1.32 (0.03) | |
| SVC | 0.90 (0.01) | 0.91 (0.01) | NA | 0.90 (0.01) | 0.90 (0.01) | 0.90 (0.01) | | 0.89 (0.02) | 0.91 (0.02) | NA | 0.82 (0.04) | 0.85 (0.03) | 0.85 (0.03) | | 1.76 (0.04) | 1.67 (0.03) | NA | 3.49 (1.23) | 1.74 (0.16) | 1.29 (0.03) | |
| **10000** | | | | | | | | | | | | | | | | | | | | | |
| NNet | 0.90 (0.01) | 0.94 (0.00) | NA | 0.90 (0.01) | 0.90 (0.01) | 0.90 (0.01) | | 0.88 (0.03) | 0.94 (0.02) | NA | 0.87 (0.03) | 0.86 (0.03) | 0.86 (0.03) | | 1.31 (0.02) | 1.43 (0.02) | NA | 1.47 (0.38) | 1.26 (0.03) | 1.22 (0.02) | |
| RFC | 0.90 (0.01) | 0.91 (0.01) | NA | 0.90 (0.01) | 0.90 (0.01) | 0.90 (0.01) | | 0.89 (0.02) | 0.91 (0.02) | NA | 0.86 (0.04) | 0.88 (0.03) | 0.87 (0.03) | | 1.83 (0.03) | 1.79 (0.02) | NA | 2.20 (0.90) | 1.49 (0.06) | 1.25 (0.02) | |
| SVC | 0.90 (0.01) | 0.91 (0.01) | NA | 0.90 (0.01) | 0.90 (0.01) | 0.90 (0.01) | | 0.89 (0.02) | 0.90 (0.02) | NA | 0.85 (0.03) | 0.86 (0.03) | 0.86 (0.03) | | 1.63 (0.03) | 1.57 (0.02) | NA | 1.85 (0.72) | 1.40 (0.06) | 1.21 (0.02) | |

Table S4: Results of experiments on CIFAR10 data. Other details are as in Table S1.

| Black box | Coverage | | | | | | | Conditional coverage | | | | | | | Size / coverage | | | | | | |
|---|---|---|---|---|---|---|---|---|---|---|---|---|---|---|---|---|---|---|---|---|---|
| | SC | CV+ | JK+ | CQC | CQC-RF | HCC | | SC | CV+ | JK+ | CQC | CQC-RF | HCC | | SC | CV+ | JK+ | CQC | CQC-RF | HCC | |
| **500** | | | | | | | | | | | | | | | | | | | | | |
| NNet | 0.91 (0.02) | 0.94 (0.01) | 0.92 (0.01) | 0.90 (0.02) | 0.90 (0.03) | 0.90 (0.02) | | 0.88 (0.03) | 0.93 (0.02) | 0.90 (0.02) | 0.77 (0.06) | 0.80 (0.06) | 0.85 (0.04) | | 8.21 (0.37) | 8.43 (0.23) | 7.99 (0.25) | 9.55 (0.16) | 9.37 (0.19) | 9.05 (0.28) | |
| RFC | 0.90 (0.02) | 0.96 (0.01) | 0.96 (0.01) | 0.91 (0.02) | 0.90 (0.03) | 0.90 (0.02) | | 0.90 (0.02) | 0.96 (0.01) | 0.96 (0.02) | 0.79 (0.06) | 0.84 (0.05) | 0.90 (0.03) | | 7.80 (0.36) | 8.23 (0.32) | 8.11 (0.34) | 9.43 (0.44) | 8.67 (0.40) | 7.67 (0.37) | |
| SVC | 0.90 (0.02) | 0.93 (0.01) | 0.91 (0.02) | 0.90 (0.03) | 0.90 (0.02) | 0.90 (0.02) | | 0.89 (0.03) | 0.92 (0.02) | 0.89 (0.03) | 0.80 (0.06) | 0.82 (0.05) | 0.87 (0.03) | | 7.92 (0.37) | 7.88 (0.32) | 7.41 (0.38) | 9.39 (0.47) | 8.75 (0.41) | 7.83 (0.33) | |
| **1000** | | | | | | | | | | | | | | | | | | | | | |
| NNet | 0.90 (0.01) | 0.96 (0.01) | 0.93 (0.01) | 0.90 (0.02) | 0.90 (0.02) | 0.90 (0.01) | | 0.88 (0.03) | 0.95 (0.01) | 0.92 (0.02) | 0.77 (0.04) | 0.82 (0.04) | 0.86 (0.03) | | 7.78 (0.25) | 8.08 (0.18) | 7.82 (0.17) | 9.47 (0.16) | 9.19 (0.15) | 8.80 (0.23) | |
| RFC | 0.90 (0.01) | 0.96 (0.01) | 0.96 (0.01) | 0.90 (0.02) | 0.90 (0.02) | 0.90 (0.01) | | 0.90 (0.02) | 0.96 (0.01) | 0.96 (0.01) | 0.80 (0.06) | 0.84 (0.04) | 0.90 (0.02) | | 7.11 (0.31) | 7.55 (0.24) | 7.47 (0.24) | 9.43 (0.45) | 8.16 (0.42) | 6.90 (0.30) | |
| SVC | 0.90 (0.01) | 0.92 (0.01) | 0.90 (0.01) | 0.90 (0.02) | 0.90 (0.02) | 0.90 (0.01) | | 0.88 (0.02) | 0.92 (0.02) | 0.89 (0.02) | 0.80 (0.05) | 0.82 (0.04) | 0.88 (0.03) | | 7.35 (0.27) | 7.18 (0.23) | 6.75 (0.24) | 9.40 (0.42) | 8.58 (0.43) | 7.22 (0.31) | |
| **5000** | | | | | | | | | | | | | | | | | | | | | |
| NNet | 0.90 (0.01) | 0.96 (0.00) | NA | 0.90 (0.01) | 0.90 (0.01) | 0.90 (0.01) | | 0.89 (0.02) | 0.96 (0.01) | NA | 0.79 (0.05) | 0.84 (0.03) | 0.86 (0.02) | | 7.09 (0.14) | 7.25 (0.12) | NA | 9.16 (0.20) | 8.40 (0.14) | 7.86 (0.19) | |
| RFC | 0.90 (0.01) | 0.95 (0.00) | NA | 0.90 (0.01) | 0.90 (0.01) | 0.90 (0.01) | | 0.89 (0.02) | 0.94 (0.01) | NA | 0.82 (0.04) | 0.86 (0.03) | 0.90 (0.02) | | 6.03 (0.15) | 6.36 (0.11) | NA | 8.81 (0.51) | 7.33 (0.29) | 5.81 (0.15) | |
| SVC | 0.90 (0.01) | 0.91 (0.01) | NA | 0.90 (0.01) | 0.90 (0.01) | 0.90 (0.01) | | 0.89 (0.02) | 0.90 (0.01) | NA | 0.81 (0.04) | 0.84 (0.03) | 0.89 (0.02) | | 6.09 (0.12) | 5.99 (0.10) | NA | 8.84 (0.41) | 7.75 (0.29) | 5.96 (0.12) | |
| **10000** | | | | | | | | | | | | | | | | | | | | | |
| NNet | 0.90 (0.01) | 0.96 (0.00) | NA | 0.90 (0.01) | 0.90 (0.01) | 0.90 (0.01) | | 0.89 (0.02) | 0.95 (0.01) | NA | 0.81 (0.05) | 0.85 (0.03) | 0.87 (0.02) | | 5.99 (0.16) | 5.89 (0.09) | NA | 8.89 (0.26) | 7.92 (0.12) | 5.93 (0.24) | |
| RFC | 0.90 (0.01) | 0.94 (0.00) | NA | 0.90 (0.01) | 0.90 (0.01) | 0.90 (0.01) | | 0.90 (0.02) | 0.94 (0.01) | NA | 0.83 (0.04) | 0.86 (0.03) | 0.89 (0.02) | | 5.69 (0.09) | 5.99 (0.05) | NA | 8.18 (0.63) | 7.13 (0.22) | 5.44 (0.09) | |
| SVC | 0.90 (0.01) | 0.91 (0.00) | NA | 0.90 (0.01) | 0.90 (0.01) | 0.90 (0.01) | | 0.89 (0.02) | 0.89 (0.02) | NA | 0.82 (0.03) | 0.84 (0.03) | 0.89 (0.02) | | 5.87 (0.08) | 5.83 (0.06) | NA | 8.37 (0.48) | 7.66 (0.18) | 5.74 (0.08) | |

12



\end{document}